\documentclass[12pt]{article}
\usepackage{epsfig}
\usepackage{float}

\date{}

\newcommand{\ba}{\begin{array}}
\newcommand{\ea}{\end{array}}
\newcommand{\bd}{\begin{displaymath}}
\newcommand{\ed}{\end{displaymath}}
\newcommand{\bi}{\begin{itemize}}
\newcommand{\ei}{\end{itemize}}
\newcommand{\benu}{\begin{enumerate}}
\newcommand{\eenu}{\end{enumerate}}
\newcommand{\be}{\begin{equation}}
\newcommand{\ee}{\end{equation}}
\newcommand{\bea}{\begin{eqnarray}}
\newcommand{\eea}{\end{eqnarray}}
\usepackage{amsfonts}
\usepackage{amssymb}
\usepackage[centertags]{amsmath}
\usepackage{graphicx}
\usepackage{hyperref}
\usepackage{booktabs}
\usepackage{theorem}
\usepackage{array}
\usepackage{epsfig}
\usepackage{colordvi}
\usepackage{color}
\usepackage{ulem}
\usepackage{cite}
\usepackage{color}
\usepackage{subfigure}
\usepackage{pdflscape}

\def\1{\mathbf{1}}
\def\3{\mathbf{3}}
\def\2{\mathbf{2}}

\def\ltap{\ \raisebox{-.4ex}{\rlap{$\sim$}} \raisebox{.4ex}{$<$}\ }
\def\gtap{\ \raisebox{-.4ex}{\rlap{$\sim$}} \raisebox{.4ex}{$>$}\ }

\newcommand{\bec}{\begin{cases}}
\newcommand{\eec}{\end{cases}}
\newcommand{\beq}{\begin{equation*}}
\newcommand{\eeq}{\end{equation*}}
\makeatletter

\newcommand{\Rmnum}[1]{\expandafter\@slowromancap\romannumeral #1@}
\makeatother

\begin{document}

\begin{titlepage}
\vspace*{-15mm}
\begin{flushright}
SISSA 13/2024/FISI\\
\end{flushright}
\vspace*{0.7cm}

\begin{center}
{\bf{\Large Neutrino 
Tomography of the Earth:\\
\vspace*{0.2cm}
the Earth Total Mass, Moment of Inertia and \\
\vspace*{0.3cm}
Hydrostatic Equilibrium Constraints
}}\\

\vspace{0.4cm} 
S. T. Petcov$\mbox{}^{a,b)}$ 
\footnote{Also at: Institute of Nuclear Research and Nuclear Energy,
Bulgarian Academy of Sciences, 1784 Sofia, Bulgaria.}  
\\[1mm]
\end{center}
\vspace*{0.50cm}
\centerline{$^{a}$ \it INFN/SISSA, Via Bonomea 265, 34136 Trieste, Italy}
\vspace*{0.2cm}
\centerline{$^{b}$ \it Kavli IPMU (WPI), University of Tokyo,
5-1-5 Kashiwanoha, 277-8583 Kashiwa, Japan}
\vspace*{0.8cm}

\begin{abstract}
\noindent 
We investigate the implications of  the constraints following from 
the precise knowledge of the total Earth mass, $M_\oplus$, 
and moment of inertia, $I_\oplus$, 
and from the requirement that Earth be in hydrostatic equilibrium (EHE), 
in the neutrino tomography studies of the Earth density structure.
In order to estimate the sensitivity of a given neutrino detector 
to possible deviations of the inner core (IC), outer core (OC), core (IC + OC) 
and mantle Earth densities from those obtained 
using geophysical and seismological data and described by
the preliminary reference Earth model  (PREM), in the statistical analyses 
performed within the neutrino tomography studies 
one typically varies the density of each of these  structures. 
These variations, however, must respect 
the  $M_\oplus$,  $I_\oplus$ and EHE constraints.
Working with PREM average densities we 
derive the $M_\oplus$, $I_\oplus$ and EHE   
constraints on the possible density variations 
when one approximates the Earth density structure 
with i) three layers - mantle, outer core and inner core, 
and ii) four layers - upper mantle, lower mantle, 
outer core and inner core. 
We get drastically different results in the two cases. 

\end{abstract}


\end{titlepage}
\setcounter{footnote}{0}

\vspace{-0.4cm}

\newpage

\section{Introduction}

At present our knowledge about the interior composition of the Earth 
and its density structure is based primarily on seismological and 
geophysical data (see, e.g., 
\cite{McDonough:2003,Bolt:1991,Kennett:1998,Masters:2003,Nimmo:2022,Koper:2022,Waszek:2023,McDonough:2023,Noe:2023}).
These data were used to construct the 
Preliminary Reference Earth Model (PREM)  \cite{PREM} of the density 
distribution of the Earth. In the PREM model, 
the Earth density distribution $\rho_{\rm E}$ is assumed to be 
spherically symmetric, $\rho_{\rm E} = \rho_{\rm E}(r)$, 
$r$ being the distance from the Earth center,
and there are three major density structures (or layers)-
the inner core, outer core and the mantle, and
a certain number of substructures (shells).
In addition there are three shells below, but very close to, 
the Earth surface - the ocean and the two 
crust ``surface'' layers.
The mantle has five shells in the model. 
The mean Earth radius is $R_{\oplus} = 6371$ km; the Earth core
has a radius of $R_c= 3480$ km, 
with the IC and OC extending respectively 
from $r = 0$ to $r = 1221.5$ km, 
and from $r = 1221.5$ km to $r = 3480$ km.
The mantle layer is located between the outer core and 
the lower crust layer, i.e., for $r$ in the interval 
$R_{\rm C} \leq r \leq R_{\rm man}$, with $R_{\rm man} = 6346.6$ km.
The mean densities of the mantle, the inner and out core are 
respectively $\bar{\rho}_{\rm man} = 4.66$ g/cm$^3$,
 $\bar{\rho}_{\rm IC} = 12.89$ g/cm$^3$ and
$\bar{\rho}_{\rm OC} = 10.90$ g/cm$^3$.
The change of density between the mantle and the outer core 
is described by a step function.

 The determination of the radial density distributions in the mantle, 
outer core and inner core (or core)
$\rho_{\rm man}(r)$, $\rho_{\rm OC}(r)$ and $\rho_{\rm IC}(r)$
(or $\rho_{\rm C}(r)$) from seismological and geophysical data is not direct 
and suffers from uncertainties 
\cite{Bolt:1991,Kennett:1998,Masters:2003,Nimmo:2022}. 
%
%
An approximate and perhaps rather conservative estimate of this uncertainty 
for $\rho_{man}(r)$ is $\sim 5\%$; for the core density  
$\rho_{c}(r)$ it is larger and can 
be significantly larger \cite{Bolt:1991,Kennett:1998,Masters:2003,Nimmo:2022}. 
It was concluded in  \cite{Masters:2003} (see also \cite{Nimmo:2022}), 
in particular, that the density increase across the inner core - outer core 
boundary is known with an uncertainty of about 20\%.

A precise knowledge of the Earth's density distribution 
and of the average densities of the Earth's three different major structures - 
the mantle, outer core and inner core -
is essential for understanding the physical conditions and fundamental 
aspects of the structure and properties of the Earth's interior
(including the dynamics of mantle and core, the bulk composition of 
the Earth's three structures, the generation, properties and evolution 
of the Earth's magnetic field and the gravity field of the Earth) 
\cite{Bolt:1991,Yoder:1995,McDonough:2003,McDonough:2008zz}.
The thermal evolution of the Earth's core, in particular,
depends critically on the density change across the inner core - 
outer core boundary (see, e.g., \cite{Baffet:1991}).

 A unique alternative method of obtaining information about 
the density profile of the Earth is  
the neutrino tomography of the Earth 
\cite{PlaZavatt1973,VolkovaZatsepin1974,Nedyalkov:1981yy,Nedyalkov:2,DeRujula:1983ya,Wilson:1983an,Askar:1984,Borisov:1986sm,Borisov:1989kh,Winter:2006vg,Kuo95,Jain:1999kp,Reynoso:2004dt,Gonzalez-Garcia:2007wfs}.
The propagation of the active flavour neutrinos 
and antineutrinos 
$\nu_\alpha$ and $\bar{\nu}_\alpha$, $\alpha=e,\mu,\tau$,
in the Earth is affected by the Earth matter.
The original idea of neutrino Earth tomography is based on 
the observation that the cross section of the neutrino-nucleon 
interaction rises with energy. For neutrinos with energies 
$E_\nu \gtrsim$ a few TeV, the inelastic scattering off protons and neutrons 
leads to absorption of neutrinos and thus to 
attenuation of the initial neutrino flux. The magnitude of the 
attenuation depends on the Earth matter density along the neutrino path.
Attenuation data for neutrinos with different path-lengths 
in the Earth carry information about the matter density distribution 
in the Earth interior. 
The absorption method of Earth tomography with accelerator 
neutrino beams, which is difficult (if not impossible) to realise 
in practice was discussed first in \cite{PlaZavatt1973,VolkovaZatsepin1974}
and later in grater detail in  
\cite{Nedyalkov:1981yy,Nedyalkov:2,DeRujula:1983ya,Wilson:1983an,Askar:1984,Borisov:1986sm,Borisov:1989kh,Winter:2006vg,Kuo95,Jain:1999kp,Reynoso:2004dt,Gonzalez-Garcia:2007wfs}.

  The oscillations between the active flavour neutrinos and antineutrinos, 
$\nu_{\alpha} \leftrightarrow \nu_{\beta}$ 
and $\bar{\nu}_{\alpha} \leftrightarrow \bar{\nu}_{\beta}$, 
$\alpha,\beta=e,\mu$ 
having energies in the range $E \sim (0.1 - 15.0)$ GeV and 
traversing the Earth can be strongly modified by the Earth matter effects 
\cite{MSW1,Barger80,MSW2,Petcov:1998su,Chizhov:1999az,AMS:2006hb}
(see also, e.g., \cite{Chizhov:1998ug,AMS:2005yj}, 
the review \cite{ParticleDataGroup:2018ovx} and references quoted therein).
These modifications depend on the Earth matter density 
along the path of the neutrinos
\footnote{More precisely, they depend on 
the electron number density $N_e(r)$,
 $N_e(r) = \rho_{rm E}(r) Y_e(r)/m_{N}$, $Y_e(r)$ and $m_N$ being 
the electron fractions number and nucleon mass.
It follows from the studies of the Earth matter composition
that $Y_e \cong (0.490-0.0496)$ in the mantle and 
$Y_e \cong (0.466-0.471)$ in the core 
(see, e.g., \cite{Bardo:2015,Kaminski:2013,Sakamaki:2009,EarthRef}).
Taking into account these uncertainties has no effect on the 
results on the densities of the mantle and the core 
obtained in neutrino tomography studies (see, e.g.,\cite{Donini:2018tsg}).
The relation between $N_e(r)$ and $\rho_{rm E}(r)$ 
can be used to obtain information about $Y_e(r)$, 
i.e., the composition of the core and mantle layers 
of the Earth (see, e.g., \cite{Rott:2015kwa,Bourret:2020zwg}). 
}. 
Thus, by studying the effects of Earth matter 
on the oscillations of, e.g., $\nu_\mu$ and $\nu_e$ ($\bar{\nu}_\mu$ and 
$\bar{\nu}_e$) neutrinos traversing the Earth along different trajectories 
it is possible to obtain information about the Earth density 
distribution. 

Atmospheric neutrinos (see, e.g., \cite{Gaisser:2002jj,Honda:2015fha}) 
are a perfect tool  for performing Earth tomography.
Consisting of  significant fluxes of 
muon and electron neutrinos and antineutrinos,
$\nu_\mu$, $\nu_e$, $\bar{\nu}_\mu$ and $\bar{\nu}_e$, 
produced in the interactions of cosmic rays with the Earth 
atmosphere, they have a wide range of energies 
spanning the interval from a few MeV to multi-GeV to multi-TeV. 
Being produced isotropically in the upper part of the Earth atmosphere 
at a height of $\sim 15$ km, they travel 
distances from $\sim 15$ km to 12742 km before reaching 
detectors located on the Earth surface, 
crossing the Earth along all possible directions 
and thus ``scanning'' the Earth interior.  
The interaction rates that allow to get 
information about the Earth density distribution 
can be obtained in the currently taking data IceCube
experiment \cite{IceCube,IceCube:Neutrino2024} and its 
planned upgrade \cite{IceCube:Neutrino2024}, in the 
ORCA \cite{KM3Net:2016zxf},   
Hyper Kamiokande \cite{Hyper-Kamiokande:2018ofw} 
and DUNE \cite{DUNE:2018tke} experiments which are under construction,
and in the planned INO \cite{ICAL:2015stm} experiment.

The idea of using the absorption method of Earth tomography with 
atmospheric neutrinos was discussed first,  
to our knowledge, in \cite{Gonzalez-Garcia:2007wfs}. 
In 2018 in \cite{Donini:2018tsg} the authors 
used the data of the IceCube experiment on multi-TeV atmospheric  
$\nu_{\mu}$ and $\bar{\nu}_\mu$ with sufficiently long paths in the Earth 
and obtained information about the Earth density distribution, which, although 
not very  precise, broadly agrees with the PREM model. 
More specifically, in \cite{Donini:2018tsg} it is assumed that 
the Earth density distribution is spherically symmetric.
The analysis is performed with a five layer Earth model: 
the inner core, two equal width layers of the outer core 
and two equal width layers of the mantle. 
The densities in each of the five layers are varied independently. 
The external constraints on the Earth total mass $M_{\oplus}$  
and moment of inertia $I_\oplus$, which are known with a remarkable 
high precision \cite{EarthMI1,EarthMI2,EarthMI3}, 
as well as the Earth hydrostatic 
equilibrium (EHE) constraint,
were not applied. The results are obtained with the IceCube 
data on the zenith angle dependence of the 
fluxes of up-going 
atmospheric $\nu_\mu$ and $\bar{\nu}_\mu$ 
with energies in the interval $E = (1.5 - 20.0)$ TeV 
\cite{IceCube:2016rnb}. Four different models of the initial fluxes 
of atmospheric $\nu_\mu$ and $\bar{\nu}_\mu$ were used in the analysis. 
The value of the Earth mass found in 
\cite{Donini:2018tsg}, $M^\nu_\oplus = (6.0 ^{+1.6}_{-1.3})\times 10^{24}$ kg, 
is in good agreement  with gravitationally determined value  
\cite{EarthMI1,EarthMI2}, 
$M_\oplus = (5.9722 \pm 0.0006) \times 10^{24}~{\rm kg}$.
%
Thus, the Earth was ``weighted'' with neutrinos.
The results obtained in \cite{Donini:2018tsg}  
contain evidence at $2\sigma$ C.L. 
that the core is denser than the mantle:
$\bar{\rho}^\nu_{c} ({\rm 3layer}) 
- \bar{\rho}^\nu_{man}({\rm 2layer}) 
= (13.1^{+5.8}_{-6.3})$ g/cm$^3$,
where $\bar{\rho}^\nu_{c} ({\rm 3layer})$ and 
$\bar{\rho}^\nu_{man}({\rm 2layer})$ are the values of the average 
core and mantle densities determined in \cite{Donini:2018tsg}. 
This was the first time 
the study of neutrinos traversing the Earth provided information 
of the Earth interior and marked the beginning of 
real experimental data driven neutrino tomography of the Earth.

The neutrino oscillation tomography of the Earth with 
IceCube, ORCA, DUNE and INO detectors 
is discussed in 
\cite{Agarwalla:2012uj,Rott:2015kwa,Winter:2015zwx,Bourret:2017tkw,Bourret:2020zwg,Kumar:2021faw,Capozzi:2021hkl,Kelly:2021jfs,Denton:2021rgt,Maderer:2022toi,Upadhyay:2022jfd,Raikwal:2023jkf,Upadhyay:2024gra} 
(the early studies include \cite{Choubey:2008_2011,Choubey:2014} 
briefly described in \cite{Capozzi:2021hkl}).
In order to estimate the sensitivity of a given neutrino detector 
to possible deviations of the inner core (IC), outer core (OC), core (IC + OC) 
and mantle Earth densities from those obtained 
using geophysical and seismological data and described by
PREM, in the statistical analyses 
performed  within the neutrino tomography studies 
one typically varies the density of each of these  structures. 
These variations, however, must respect 
the  $M_\oplus$,  $I_\oplus$ and EHE constraints.
In the studies performed so far only in \cite{Kelly:2021jfs} 
all three  $M_\oplus$,  $I_\oplus$ and EHE constraints 
were taken into account. The  $M_\oplus$ and EHE constraints, 
but not the  $I_\oplus$ one, were implemented in the analyses performed 
in \cite{Capozzi:2021hkl,Raikwal:2023jkf}.  

 The existing studies provide very limited 
information about the implications, and thus on the relevance, if any,
 of the three external constraints -  $M_\oplus$,  $I_\oplus$ and EHE -
in the neutrino tomography studies of the Earth density distribution.
The aim of the present letter is to assess the importance of 
the three constraints in the neutrino tomography 
studies of the Earth interior. Working with PREM average densities 
we derive the $M_\oplus$, $I_\oplus$ and EHE   
constraints on the possible density variations 
in two illustrative and widely considered cases, namely,  
when one approximates the Earth density structure 
with i) three layers - mantle, outer core and inner core, 
and ii) four layers - upper mantle, lower mantle, 
outer core and inner core. 

 The article is organised as follows. In Section 2 
we derive the contributions of the ``surface'' layers, 
the mantle, outer core and inner core 
to the Earth total mass and moment of inertia 
in terms of average densities of these layers 
calculated using the corresponding 
PREM density distributions. In Section 3 
we implement the   $M_\oplus$, $I_\oplus$ and EHE   
constraints considering uniform ($r-$independent) 
deviations of the average densities of the Earth layers 
from their respective PREM values. In sub-sections 3.1 and 3.2
We consider the cases when  one approximates the
Earth density structure with 
i) three layers - mantle, outer core and inner core, and ii)
four layers - upper mantle, lower mantle, outer core and inner core.  
We obtain the equations which the deviation parameters should satisfy and 
provide the analytic solutions of these equations.
We further comment on the implications of the results obtained.
Section 4 contains a brief summary.

\vspace{0.3cm}
%
\section{Earth Total Mass and Moment of Inertia:
the Mantle, Outer Core, Inner Core and ``Surface'' 
Layers Contributions 
}
\label{sec:MOCICtoMI}
%

\vspace{0.3cm}
In order to estimate the sensitivity of any detector 
to the IC, OC, core (IC + OC) and mantle densities, 
one typically varies the density 
in one of these structures - the structure of interest.
These variations, however, must respect the constraints 
following form the precise knowledge of the total Earth mass,
$M_\oplus$, and momentum of inertia, $I_\oplus$ 
\cite{EarthMI1,EarthMI2,EarthMI3}:
\begin{eqnarray}
\label{eq:ME}
& M_\oplus =  (5.9722 \pm 0.0006)\times 10^{24}~{\rm kg}\,,\\[0.25cm]
\label{eq:IE}
&  I_\oplus = (8.01736 \pm 0.00097)\times 10^{37}~{\rm kg\, m^2}\,.
\end{eqnarray}
%
The mean value of $I_\oplus$ can be related to the Earth mean total mass and 
radius $R_\oplus$:
\begin{equation}
I_\oplus = 
C_{\rm G(PREM)} \,M_\oplus\,R^2_\oplus\,.
\label{eq:IMR2}
\end{equation}
%
where $C_{\rm G}$  ($C_{\rm PREM}$) 
is a constant determined in gravity experiments 
(the PREM model):
\begin{equation}
C_{\rm G} = 0.330745\,,~~~C_{\rm PREM} = 0.33080\,.
\label{eq:C}
\end{equation}
%

 One can implement the total Earth mass and moment of inertia constraints 
by compensating the variation of the density in the structure of interest 
by corresponding change of the densities in at least two of the  
other structures. 

 Following PREM, we will assume that the Earth density distribution 
$\rho_{\rm E}$ is spherically symmetric, $\rho_{\rm E} = \rho_{\rm E}(r)$, 
where $r$ is the distance from the Earth center. 
The Earth density structure consists of inner core, outer core, 
mantle and three relatively thin layers close to the Earth surface - 
the ocean and two crust ``surface'' layers, 
having masses $M_{\rm IC}$, $M_{\rm OC}$, $M_{\rm man}$ and
$M_{\rm o+c1+c2}$, and moments of inertia 
$I_{\rm IC}$, $I_{\rm OC}$, $I_{\rm man}$ and
$I_{\rm o+c1+c2}$, respectively.
Correspondingly, the total Earth mass in PREM is given by:
\begin{eqnarray}
& M_\oplus = M_{\rm IC} + M_{\rm OC} + M_{\rm man} + M_{\rm o+c1+c2} = 
\int^{R_\oplus}_{0} 4\pi \rho_{\rm E}(r) r^2 dr 
\nonumber
\\[0.30cm] 
&= 4\pi \left [ 
\int^{R_{\rm IC}}_{0} \rho_{\rm IC}(r) r^2 dr + 
\int^{R_{\rm C}}_{R_{\rm IC}} \rho_{\rm OC}(r) r^2 dr +
\int^{R_{\rm man}}_{R_{\rm C}} \rho_{\rm man}(r) r^2 dr 
+ \int^{R_{\oplus}}_{R_{\rm man}} \rho_{\rm o+c1+c2}(r) r^2 dr \right]\,
\label{eq:ME2}
\end{eqnarray}
%
where  $\rho_{\rm IC}(r)$,  $\rho_{\rm OC}(r)$, $\rho_{\rm mant}(r)$
and $\rho_{\rm o+c1+c2}(r)$
are the inner core (IC), outer core 
(OC), the mantle and the ``surface'' layers
densities as a function of $r$ and, 
according to PREM,
\begin{equation} 
R_{\rm IC} = 1221.5~{\rm km}\,,~R_{\rm OC} = R_{\rm C} = 3480~{\rm km}\,,
~R_{\rm man} = 6346.6~{\rm km}\,,~R_\oplus = 6371~{\rm km}\,. 
\label{eq:RICCE}
\end{equation}~
%
Note that $R_{\rm man}$ is smaller than the mean radius of the 
Earth  $R_\oplus$ due to the presence in PREM of 
an ``ocean'' and two crust ``surface'' layers. 

Similarly, the Earth total moment of inertia can be written as:
\begin{eqnarray}
& I_\oplus = I_{\rm IC} + I_{\rm OC} + I_{\rm man} +
I_{\rm o+c1+c2} = \int^{R_\oplus}_{0} \dfrac{8\pi}{3} \rho_{\rm E}(r) r^4 dr
\nonumber
\\[0.30cm] 
& = \dfrac{8\pi}{3} \left [ 
\int^{R_{\rm IC}}_{0} \rho_{\rm IC}(r) r^4 dr + 
\int^{R_{\rm C}}_{R_{\rm IC}} \rho_{\rm OC}(r) r^4 dr +
\int^{R_{\rm man}}_{R_{\rm C}} \rho_{\rm man}(r) r^4 dr 
+ \int^{R_{\oplus}}_{R_{\rm man}} \rho_{\rm o+c1+c2}(r) r^4 dr
\right ]\,.
\label{eq:IE2}
\end{eqnarray}
%

 It follows from the seismological and geophysical data 
that in order for Earth to be in hydrostatic equilibrium  
the following inequalities should always hold:
\begin{equation}
\rho_{man} < \rho_{OC} <  \rho_{IC}\,.
\label{eq:equil}
\end{equation}
%
We will refer to these inequalities as Earth hydrostatic equilibrium (EHE) 
constraint. The EHE constraint should also be implemented in the 
studies of the Earth  density structure using the method of neutrino tomography.

In our further analysis we will use the the IC, OC, mantle
and the ``surface'' layers  density 
distributions given by PREM. In PREM, the mantle is represented by 
five layers with different distributions in each layer. 
In addition, as we have already indicated, 
there are three thin layers with constant density 
close to the Earth's surface: one ``ocean'' and two crust 
``surface'' layers.

In PREM the IC and OC density distributions have the form:
\begin{eqnarray}
\label{eq:rhoIC}
& \rho_{\rm IC}(r) = 13.0885 - 8.8381 \dfrac{r^2}{R^2_\oplus}\,,~
~r=(0-1221.5)~{\rm km}\,;\\[0.30cm]
& \rho_{\rm OC}(r)  = 12.5815 - 1.2638\dfrac{r}{R_\oplus}
- 3.6426\dfrac{r^2}{R^2_\oplus} - 5.5281\dfrac{r^3}{R^3_\oplus}\,,~
r=(1221.5-3480.0)~{\rm km}\,,
\label{eq:rhoOC}
\end{eqnarray}
%
where $\rho_{\rm IC}(r)$ and $\rho_{\rm OC}(r)$
are  in units of g/cm$^3$ and we have given also 
for convenience the ``widths'' of the IC and OC layers.
Similarly, the densities (in units of  g/cm$^3$)
and the widths of the ocean, the two crust 
and the five mantle layers read \cite{PREM}: 
\begin{eqnarray}
\label{eq:rhoocean}
& \rho_{\rm ocean}(r) = 1.020\,,
~r=(6368-6371)~{\rm km}\,;\\[0.30cm]
\nonumber
& \rho_{\rm crust1}(r) = 2.600\,,~r=(6356-6368)~{\rm km}\,;\\[0.30cm]
\label{eq:rhocrust}
& \rho_{\rm crust2}(r) = 2.900\,,~r=(6346.6-6356.0)~{\rm km}\,;\\[0.30cm]
\nonumber
& \rho_{\rm m1}(r)  = 2.6910 + 0.6924\dfrac{r}{R_\oplus}\,,
~r=(6151-6346.6)~{\rm km}\,;\\[0.30cm]
\nonumber
& \rho_{\rm m2}(r)  = 7.1089 - 3.8045\dfrac{r}{R_\oplus}\,,
~r=(5971-6151)~{\rm km}\,;\\[0.30cm]
\nonumber
& \rho_{\rm m3}(r)  = 11.2494 - 8.0298\dfrac{r}{R_\oplus} \,,
~r=(5771-5971)~{\rm km}\,;\\[0.30cm]
\nonumber
& \rho_{\rm m4}(r)  = 5.3197 - 1.4836\dfrac{r}{R_\oplus}\,,
~r=(5701-5771)~{\rm km}\,;\\[0.30cm]
& \rho_{\rm m5}(r)  = 7.9565 - 6.4761\dfrac{r}{R_\oplus}
+ 5.5283\dfrac{r^2}{R^2_\oplus} - 3.0807\dfrac{r^3}{R^3_\oplus}\,,
~r=(3480-5701)~{\rm km}\,.
\label{eq:rhoman}
\end{eqnarray}
%

Using Eqs. (\ref{eq:ME2}) and (\ref{eq:IE2}) - (\ref{eq:rhoman}), 
it is not difficult to obtain the contributions of the IC, OC, 
the mantle and ``surface'' layers 
in $M_\oplus$ and $I_\oplus$. For the contributions of the 
IC and OC layers, $M_{\rm IC}$, $I_{\rm IC}$ and  $M_{\rm OC}$, $I_{\rm OC}$,
we get:
\begin{eqnarray}
\label{eq:MIC}
& M_{\rm IC} = \dfrac{4\pi}{3}\,\bar{\rho}_{\rm IC}\,R^3_{\rm IC}
\,,\\[0.30cm]
\label{eq:IIC}
& I_{\rm IC} = \dfrac{2}{5}\,\dfrac{4\pi}{3}\,
\bar{\rho}^{\rm I}_{\rm IC}\,R^5_{\rm IC} = 
\dfrac{2}{5}\,\dfrac{\bar{\rho}^{\rm I}_{\rm IC}}{\bar{\rho}_{\rm IC}}\,
M_{\rm IC}\,R^2_{\rm IC}
\,,\\[0.30cm]
\label{eq:MOC}
& M_{\rm OC} = \dfrac{4\pi}{3}\,\bar{\rho}_{\rm OC}\,
\left (R^3_{\rm C} - R^3_{\rm IC}\right)
\,,~~R_{\rm C}=R_{\rm OC}\,,\\[0.30cm]
\label{eq:IOC}
& I_{\rm OC} = \dfrac{2}{5}\,\dfrac{4\pi}{3}\,\bar{\rho}^{\rm I}_{\rm OC}
\left (R^5_{\rm C} -  R^5_{\rm IC} \right)
=  \dfrac{2}{5}\,\dfrac{\bar{\rho}^{\rm I}_{\rm OC}}{\bar{\rho}_{\rm OC}}\,
 M_{\rm OC}\, \dfrac{R^5_{\rm C} -  R^5_{\rm IC}}{R^3_{\rm C} - R^3_{\rm IC}}\,,
\end{eqnarray}
%
where $\bar{\rho}_{\rm IC}$, $\bar{\rho}^{\rm I}_{\rm IC}$, 
$\bar{\rho}_{\rm OC}$ and $\bar{\rho}^{\rm I}_{\rm OC}$ 
are average densities,
\begin{eqnarray}
\label{eq:rhoIC2}
&\bar{\rho}_{\rm IC} = \dfrac{4\pi \int^{R_{\rm IC}}_{0} \rho_{\rm IC}(r) r^2 dr}
{4\pi \int^{R_{\rm IC}}_{0} r^2 dr}
= 13.0885 - 8.8381\, \dfrac{3R^2_{\rm IO}}{5R^2_\oplus} 
\,, \\[0.30cm]
\label{eq:rhoIIC}
& \bar{\rho}^{\rm I}_{\rm IC} =
\dfrac{\frac{8\pi}{3}\int^{R_{\rm IC}}_{0} \rho_{\rm IC}(r) r^4 dr}
{\frac{8\pi}{3}\int^{R_{\rm IC}}_{0} r^4 dr}
=  13.0885 - 8.8381\,\dfrac{5R^2_{\rm IC}}{7R^2_\oplus} 
\,,\\[0.30cm]
\label{eq:rhoOC2}
& \bar{\rho}_{\rm OC} = \dfrac{4\pi \int^{R_{\rm C}}_{R_{\rm IC}}\rho_{\rm OC}(r) r^2 dr}
{4\pi \int^{R_{\rm C}}_{R_{\rm IC}} r^2 dr}
\nonumber
\\[0.30cm]
& =  12.5815 - 1.2638\,\dfrac{3(R^4_{\rm OC} - R^4_{\rm IO})}
{4R_\oplus (R^3_{\rm OC}-R^3_{\rm IO})}- 3.6426\,\dfrac{3(R^5_{\rm OC}-R^5_{\rm IO})}
{5R^2_\oplus (R^3_{\rm OC}-R^3_{\rm IO})} 
- 5.5281\,\dfrac{3(R^6_{\rm OC}-R^6_{\rm IO})}{6R^3_\oplus (R^3_{\rm OC}-R^3_{\rm IO})} 
\,,\\[0.30cm]
\label{eq:rhoIOC}
& \bar{\rho}^{\rm I}_{\rm OC} = 
 \dfrac{\frac{8\pi}{3}\int^{R_{\rm C}}_{R_{\rm IC}}\rho_{\rm OC}(r) r^4 dr}
{\frac{8\pi}{3} \int^{R_{\rm C}}_{R_{\rm IC}} r^4 dr}
\nonumber
\\[0.30cm]
& = 12.5815 - 1.2638\,
\dfrac{5(R^6_{\rm OC}-R^6_{\rm IO})}{6R_\oplus (R^5_{\rm OC}-R^5_{\rm IO})}
- 3.6426\,\dfrac{5(R^7_{\rm OC}-R^7_{\rm IO})}{7R^2_\oplus (R^5_{\rm OC}-R^5_{\rm IO})} 
- 5.5281\,\dfrac{5(R^8_{\rm OC}-R^8_{\rm IO}) }
{8R^3_\oplus (R^5_{\rm OC}-R^5_{\rm IO})}\,.
\end{eqnarray}
%
Using the values of $R_{\rm IC}$, $R_{\rm OC}$ and  $R_\oplus$
we find:
\begin{eqnarray}
\label{eq:rhoICIIC}
&\bar{\rho}_{\rm IC} = 12.8936\,\dfrac{g}{cm^3}\,,~~
\bar{\rho}^{\rm I}_{\rm IC} =12.8564\,\dfrac{g}{cm^3}
= 0.9971\,\bar{\rho}_{\rm IC}\,,\\[0.30cm] 
& \bar{\rho}_{\rm OC} = 10.90070\,\dfrac{g}{cm^3}\,,~~
\bar{\rho}^{\rm I}_{\rm OC} = 10.6583\,\dfrac{g}{cm^3}
= 0.9778\,\bar{\rho}_{\rm OC}\,.
\label{eq:rhoOCIOC}
\end{eqnarray}
%
Note that $\bar{\rho}^{\rm I}_{\rm IC}$ and $\bar{\rho}^{\rm I}_{\rm OC}$ 
differ respectively from $\bar{\rho}_{\rm IC}$ 
and $\bar{\rho}_{\rm OC}$, being somewhat smaller.

We can consider also the the total mass and moment of inertia of the 
Earth core:
\begin{eqnarray}
\label{eq:MC}
& M_{\rm C} = M_{\rm IC} + M_{\rm OC} = 
\dfrac{4\pi}{3}\,\bar{\rho}_{\rm C}\,R^3_{\rm C}\,,\\[0.30cm]
\label{eq:IC}
& I_{\rm C} = I_{\rm IC} + I_{\rm OC} = 
\dfrac{2}{5}\,\dfrac{4\pi}{3}\,\bar{\rho}^{\rm I}_{\rm C}\,R^5_{\rm C}
= \dfrac{2}{5}\,\dfrac{\bar{\rho}^{\rm I}_{\rm C}}{\bar{\rho}_{\rm C}}\,
 M_{\rm C}\,R^2_{\rm C}\,,
\end{eqnarray}
%
where
\begin{eqnarray}
\label{eq:rhoC}
& \bar{\rho}_{\rm C} = \bar{\rho}_{\rm IC}\,\dfrac{R^3_{\rm IC}}{R^3_{\rm C}} 
+ \bar{\rho}_{\rm OC}\,\left( 1 - \dfrac{R^3_{\rm IC}}{R^3_{\rm C}}\right ) 
= 10.9869\,\dfrac{g}{cm^3}\,,\\[0.30cm] 
\label{eq:rhoCI}
& \bar{\rho}^{\rm I}_{\rm C} = 
 \bar{\rho}^{\rm I}_{\rm IC}\,\dfrac{R^5_{\rm IC}}{R^5_{\rm C}} + 
\bar{\rho}^{\rm I}_{\rm OC}\,\left ( 1 - \dfrac{R^5_{\rm IC}}{R^5_{\rm C}}\right) 
 = 10.6700\,\dfrac{g}{cm^3} = 0.9712\,\bar{\rho}_{\rm C}\,.
\end{eqnarray}
%

 It is not difficult to calculate in a similar way the 
contributions of the 
five mantle layers to $M_{\oplus}$ and $I_\oplus$. 
They can be written in the form:
\begin{eqnarray}
\label{eq:Mman}
& M_{\rm man} =  \dfrac{4\pi}{3}\,\bar{\rho}_{\rm man}\,
\left (R^3_{\rm man} - R^3_{\rm C}\right )\,,~
\bar{\rho}_{\rm man} = 
 \dfrac{4\pi \int^{R_{\rm man}}_{R_{\rm C}} \rho_{\rm man}(r) r^2 dr}
{4\pi \int^{R_{\rm man}}_{R_{\rm C}} r^2 dr}
\\[0.30cm]
\label{eq:Iman}
& I_{\rm man} = \dfrac{8\pi}{15}\,\bar{\rho}^{\rm I}_{\rm man}\, 
 \left(R^5_{\rm man} -  R^5_{\rm C}\right) 
=  \dfrac{2}{5}\,\dfrac{\bar{\rho}^{\rm I}_{\rm man}}{\bar{\rho}_{\rm man}}\,
M_{\rm man}\, \dfrac{R^5_{\rm man} -  R^5_{\rm C}}{R^3_{\rm man} - R^3_{\rm C}}\,,~
\bar{\rho}^{\rm I}_{\rm man} = 
\dfrac{ \dfrac{8\pi}{3} \int^{R_{\rm man}}_{R_{\rm C}} \rho_{\rm man}(r) r^4 dr}
{ \dfrac{8\pi}{3} \int^{R_{\rm man}}_{R_{\rm C}} r^4 dr}\,.
\end{eqnarray}
%
For the average densities $\bar{\rho}_{\rm man}$ and $\bar{\rho}^{\rm I}_{\rm man}$
we get:
\begin{equation}
\label{eq:rhomIm}
\bar{\rho}_{\rm man} = 4.6651\,\dfrac{g}{cm^3}\,,~~
\bar{\rho}^{\rm I}_{\rm man} = 4.2833\,\dfrac{g}{cm^3} 
= 0.9182\,\bar{\rho}_{\rm man}\,. 
\end{equation}
%
 We remark that if we include the ``surface'' layers in the mantle, 
integrating from $R_C = 3480$ km to $R_\oplus = 6371$ km we obtain:
$\bar{\rho}_{\rm man} = 4.4551\,{\rm g/cm^3}$,
$ \bar{\rho}^{\rm I}_{\rm man} = 4.2481\,{\rm g/cm^3} 
 = 0.9535\,\bar{\rho}_{\rm man}$.

All results derived above are within the PREM framework.
For the Earth total mass and moment on inertia in 
this framework we have:  $M_\oplus =  
M_{o+c1+c2} + M_{\rm man} + M_{\rm C}$ and 
$I_\oplus = I_{o+c1+c2} + I_{\rm man} + I_{\rm C}$, 
where $M_{\rm C} = M_{\rm IC} +  M_{\rm OC}$, 
$I_{\rm C} = I_{\rm IC} +  I_{\rm OC}$
and $M_{o+c1+c2}$ and $I_{o+c1+c2}$ 
are the contributions from the ocean and 
the two crust ``surface'' layers.
Using the PREM results shown in Eqs. (\ref{eq:rhoIC}) - (\ref{eq:rhoman}), 
we find $M_{o+c1+c2} \ll M_{\rm man} + M_{\rm C}$ 
and $I_{o+c1+c2} \ll I_{\rm man} + I_{\rm C}$:
$M_{o+c1+c2}/ (M_{\rm man} + M_{\rm C}) \cong 5\times 10^{-3}$
and  $I_{o+c1+c2}/(I_{\rm man} + I_{\rm C}) \cong 0.01060$. 

We note that the ``surface'' layers 
play subleading (if not negligible) role in neutrino tomography of the 
Earth due to their relatively small widths and 
contributions to $M_\oplus$ and  $I_\oplus$. 
As far as $M_{o+c1+c2}$ and $I_{o+c1+c2}$ are kept 
constant, they do not play any role in our analysis 
and the results we will obtain are independent of 
the values of $M_{o+c1+c2}$ and $I_{o+c1+c2}$, 
as it will become clear further. 
We will keep them for consistency. 
For this purpose we introduce their average densities 
$\bar{\rho}_{\rm oc1c2}$ and $\bar{\rho}^{\rm I}_{\rm oc1c2}$:
\begin{eqnarray}
& M_{\rm o+c1+c2} =  \dfrac{4\pi}{3}\,\bar{\rho}_{\rm oc1c2}\,
\left (R^3_{\oplus} - R^3_{\rm man}\right )\,,
\nonumber
\\[0.30cm]
\label{eq:rhooc1c2}
& \bar{\rho}_{\rm oc1c2} = 
 \dfrac{4\pi \left[ 
\int^{R_{\oplus}}_{R_{\rm ocean}} \rho_{\rm ocean }(r) r^2 dr
+ \int^{R_{\rm ocean}}_{R_{\rm crust1}} \rho_{\rm crust1 }(r) r^2 dr
+ \int^{R_{\rm crust1}}_{R_{\rm crust2}} \rho_{\rm crust2 }(r) r^2 dr
\right]}
{4\pi \int^{R_{\oplus}}_{R_{\rm man}} r^2 dr}\,,
\\[0.30cm]
& I_{\rm o+c1+c2} = \dfrac{8\pi}{15}\,\bar{\rho}^{\rm I}_{\rm oc1c2}\, 
 \left(R^5_{\oplus} -  R^5_{\rm man}\right)\,,
\nonumber
\\[0.30cm]
\label{eq:rhoIoc1c2}
& \bar{\rho}^{\rm I}_{\rm oc1c2} = 
\dfrac{ \dfrac{8\pi}{3} \left[ 
\int^{R_{\oplus}}_{R_{\rm ocean}} \rho_{\rm ocean }(r) r^4 dr
+ \int^{R_{\rm ocean}}_{R_{\rm crust1}} \rho_{\rm crust1 }(r) r^4 dr
+ \int^{R_{\rm crust1}}_{R_{\rm crust2}} \rho_{\rm crust2 }(r) r^4 dr
\right]}
{\dfrac{8\pi}{3} \int^{R_{\oplus}}_{R_{\rm man}} r^4 dr}\,,
\end{eqnarray}
%
where $R_{\rm ocean} = 6368$ km, $R_{\rm crust1} = 6356$ km and 
$R_{\rm crust2} \equiv R_{\rm man} = 6346.6$ km.
For $\bar{\rho}_{\rm oc1c2}$ and 
$\bar{\rho}^{\rm I}_{\rm oc1c2}$ we get:
\begin{equation}
\bar{\rho}_{\rm oc1c2} = 2.5204\,\dfrac{g}{cm^3}\,, \quad
\bar{\rho}^{\rm I}_{\rm oc1c2} = 2.5194\,\dfrac{g}{cm^3} =
0.9996\,\bar{\rho}_{\rm oc1c2}\,.
\label{eq:rhorhoIoc1c2}
\end{equation}
%

Summarising the results obtained so far we have:
\begin{eqnarray}
& M_\oplus = M_{\rm man} + M_{\rm OC} + M_{\rm IC} + M_{o+c1+c2} 
\nonumber
\\[0.30cm]
\label{eq:MEP}
& = \dfrac{4\pi}{3}\,R^3_\oplus 
\left [\bar{\rho}_{\rm man}\,
\dfrac{R^3_{\rm man}}{R^3_\oplus}\left (1 - \dfrac{R^3_{\rm C}}{R^3_{\rm man}} \right)
+ \bar{\rho}_{\rm OC}\, \dfrac{R^3_{\rm C}}{R^3_\oplus}\,
\left (1 - \dfrac{R^3_{\rm IC}}{R^3_{\rm OC}} \right)
+ \bar{\rho}_{\rm IC}\,\dfrac{R^3_{\rm IC}}{R^3_\oplus} 
+ \bar{\rho}_{\rm oc1c2}\,\left (1 - \dfrac{R^3_{\rm man}}{R^3_{\oplus}}\right)
\right]\,,
\\[0.30cm]
& I_\oplus  =  I_{\rm man} + I_{\rm OC} + I_{\rm IC} + I_{o+c1+c2}
\nonumber
\\[0.30cm]
\label{eq:IEP} 
& = \dfrac{2}{5}\,\dfrac{4\pi}{3}\,R^5_\oplus 
\left [\bar{\rho}^{\rm I}_{\rm man}\,\dfrac{R^5_{\rm man}}{R^5_\oplus} 
\left(1 - \dfrac{R^5_{\rm C}}{R^5_{\rm man}} \right) 
+ \bar{\rho}^{\rm I}_{\rm OC}\dfrac{R^5_{\rm C}}{R^5_\oplus}
\left (1 - \dfrac{R^5_{\rm IC}}{R^5_{\rm OC}} \right)
+ \bar{\rho}^{\rm I}_{\rm IC}\dfrac{R^5_{\rm IC}}{R^5_\oplus}
+ \bar{\rho}^{\rm I}_{\rm oc1c2}\left (1 - \dfrac{R^5_{\rm man}}{R^5_{\oplus}}\right)
\,\right]\,. 
\end{eqnarray}
%
In terms of mantle and total core densities the expressions for
$M_\oplus$ and $I_\oplus$ read:
\begin{eqnarray}
& M_\oplus = M_{\rm man} + M_{\rm C} + M_{o+c1+c2} 
\nonumber
\\[0.30cm]
\label{eq:MEP2}
& = \dfrac{4\pi}{3}\,R^3_\oplus 
\left [\bar{\rho}_{\rm man}\,\dfrac{R^3_{\rm man}}{R^3_\oplus}
\left (1 - \dfrac{R^3_{\rm C}}{R^3_{\rm man}} \right)
+ \bar{\rho}_{\rm C}\, \dfrac{R^3_{\rm C}}{R^3_\oplus}
+ \bar{\rho}_{\rm oc1c2}\left (1 - \dfrac{R^3_{\rm man}}{R^3_{\oplus}}\right)
\right]\,, 
\\[0.30cm]
& I_\oplus = I_{\rm man} + I_{\rm C} + I_{o+c1+c2} 
\nonumber
\\[0.30cm]
\label{eq:IEPII}
& = \dfrac{2}{5}\,\dfrac{4\pi}{3}\,R^5_\oplus 
\left [\bar{\rho}^{\rm I}_{\rm man}\,\dfrac{R^5_{\rm man}}{R^5_\oplus} 
\left(1 - \dfrac{R^5_{\rm C}}{R^5_{\rm man}} \right) 
+ \bar{\rho}^{\rm I}_{\rm C}\dfrac{R^5_{\rm C}}{R^5_\oplus}
+ \bar{\rho}^{\rm I}_{\rm oc1c2}\left (1 - \dfrac{R^5_{\rm man}}{R^5_{\oplus}}\right)
\right]\,, 
\end{eqnarray}
%
where we have used Eqs. (\ref{eq:MC}) - (\ref{eq:rhoCI}) and 
(\ref{eq:rhooc1c2}) - (\ref{eq:rhoIoc1c2}). 
%

It proves convenient to express the constants 
$\bar{\rho}^{({\rm I})}_i$ in Eqs. (\ref{eq:IEP}) 
and (\ref{eq:IEPII}), $i={\rm IC,OC,C,man,oc1c2}$, 
as $\bar{\rho}^{({\rm I})}_i = r_i\bar{\rho}_i$.
For the ratios $r_i$ we have: 
\begin{eqnarray}
& r_{\rm man} = \dfrac{\bar{\rho}^{\rm I}_{\rm man}}{\bar{\rho}_{\rm man}} = 0.9182\,,
r_{\rm OC} = \dfrac{\bar{\rho}^{\rm I}_{\rm OC}}{\bar{\rho}_{\rm OC}} = 0.9778\,,~
r_{\rm IC} =\dfrac{\bar{\rho}^{\rm I}_{\rm IC}}{\bar{\rho}_{\rm IC}} = 0.9971\,,~
\nonumber
\\[0.20cm]
& r_{\rm C} = \dfrac{\bar{\rho}^{\rm I}_{\rm C}}{\bar{\rho}_{\rm C}} = 0.9712\,,~
r_{\rm oc1c2} 
 = \dfrac{\bar{\rho}^{\rm I}_{\rm oc1c2}}{\bar{\rho}_{\rm oc1c2}} = 0.9996\,.
\label{eq:ri}
\end{eqnarray}
%
The values of these ratios are given in 
Eqs. (\ref{eq:rhoICIIC}),  (\ref{eq:rhoOCIOC}), (\ref{eq:rhoCI}),
(\ref{eq:rhomIm}) and (\ref{eq:rhorhoIoc1c2}).

It also proves convenient to introduce the average Earth density:
\begin{eqnarray}
& \bar{\rho}_{\oplus} \equiv \dfrac{3}{4\pi} \dfrac{M_\oplus}{R^3_\oplus} 
= a_{\rm man}\,\bar{\rho}_{\rm man} + a_{\rm OC}\,\bar{\rho}_{\rm OC} 
+ a_{\rm IC}\,\bar{\rho}_{\rm IC} 
+ a_{\rm oc1c2}\,\bar{\rho}_{\rm oc1c2}
\label{eq:rhoEII}
\\[0.20cm]
& =  a_{\rm man}\,\bar{\rho}_{\rm man} + a_{\rm C}\,\bar{\rho}_{\rm C} 
+ a_{\rm oc1c2}\,\bar{\rho}_{\rm oc1c2}\,.
\label{eq:rhoEIII}
\end{eqnarray}
%
where
\begin{eqnarray}
& a_{\rm man} = \dfrac{R^3_{\rm man}}{R^3_\oplus} 
\left(1 - \dfrac{R^3_{\rm C}}{R^3_{\rm man}}\right) \cong 0.8256\,,~ 
a_{\rm OC} = \dfrac{R^3_{\rm C}}{R^3_\oplus} 
\left(1 - \dfrac{R^3_{\rm IC}}{R^3_{\rm C}} \right) 
\cong 0.15592\,,
~a_{\rm C} = \dfrac{R^3_{\rm C}}{R^3_\oplus}\cong 0.1639\,, 
\nonumber
 \\[0.20cm] 
& a_{\rm IC} = \dfrac{R^3_{\rm IC}}{R^3_\oplus} \cong 7.0479\times 10^{-3}\,.~
a_{\rm oc1c2} = \left (1 - \dfrac{R^3_{\rm man}}{R^3_{\oplus}}\right) 
\cong 0.0114\,.
\label{eq:a}
\end{eqnarray}
%

Replacing  $\bar{\rho}^{({\rm I})}_i$
with $r_i\bar{\rho}_i$ in Eqs. (\ref{eq:IEP}) 
and (\ref{eq:IEPII}), $i={\rm IC,OC,C,man,oc1c2}$, 
and using Eq. (\ref{eq:rhoEII}) 
 we get $I_\oplus = C\,M_\oplus\,R^2_\oplus$, 
which implies:
\begin{eqnarray}
& \dfrac{I_\oplus}{R^{5}_\oplus}\dfrac{5}{2}\,\dfrac{3}{4\pi} =  
C^\prime\,\bar{\rho}_\oplus = 
b_{\rm man}\,\bar{\rho}_{\rm man} + b_{\rm OC}\,\bar{\rho}_{\rm OC} 
+ b_{\rm IC}\,\bar{\rho}_{\rm IC}
+ b_{\rm oc1c2}\,\bar{\rho}_{\rm oc1c2}
\label{eq:IvsrhoE}
\\[0.25cm]
& = b_{\rm man}\,\bar{\rho}_{\rm man} + b_{\rm C}\,\bar{\rho}_{\rm C} 
+ b_{\rm oc1c2}\,\bar{\rho}_{\rm oc1c2}\,
\label{eq:IvsrhoEII}
\end{eqnarray}
%
where $C^\prime$ is a constant,  $C^\prime = 5C/2 \cong 0.8270$ and
\begin{eqnarray}
& b_{\rm man}=r_{\rm man} \dfrac{R^5_{\rm man}}{R^5_\oplus} 
\left(1 - \dfrac{R^5_{\rm C}}{R^3_{\rm man}}\right)\,,~
b_{\rm OC} = r_{\rm OC}\,\dfrac{R^5_{\rm C}}{R^5_\oplus} 
\left(1 - \dfrac{R^5_{\rm IC}}{R^5_{\rm C}}\right)\,,~
b_{\rm C} = r_{\rm C}\,\dfrac{R^5_{\rm C}}{R^5_\oplus}\,, 
\nonumber 
\\[0.25cm] 
& b_{\rm IC} = r_{\rm IC}\dfrac{R^5_{\rm IC}}{R^5_\oplus}\,,~~
b_{\rm oc1c2} =  r_{\rm oc1c2}\,\left (1 - \dfrac{R^5_{\rm man}}{R^5_{\oplus}}\right)\,.
\label{eq:b}
\end{eqnarray}
%
The constant parameters $r_{\rm man}$,  $r_{\rm OC}$,  $r_{\rm IC}$ and 
$r_{\rm oc1c2}$ are defined and their values are given in Eq. (\ref{eq:ri}).  
They are all close to, but different from, 1.  
The numerical values of $b_{\rm man}$, $b_{\rm OC}$, $b_{\rm C}$, $b_{\rm IC}$ 
and $b_{\rm oc1c2}$ read:
\begin{equation}
b_{\rm man}\cong 0.8561\,,
~b_{\rm OC}\cong  0.0473\,,~b_{\rm C}\cong 0.0472\,,~
b_{\rm IC}\cong 2.58\times 10^{-4}\,,~
b_{\rm oc1c2} \cong 0.0190\,.
\label{eq:bmOCICnum}
\end{equation}
%

The left-hand sides of Eqs. (\ref{eq:rhoEII}), (\ref{eq:rhoEIII}),  
(\ref{eq:IvsrhoE}) and  
 (\ref{eq:IvsrhoEII}) are fixed since $M_\oplus$, $I_\oplus$ and $R_\oplus$ 
are known with relatively high precision.
Thus, within the aspects of neutrino tomography of the Earth we are 
considering, Eqs. (\ref{eq:rhoEII}), (\ref{eq:rhoEIII}) and
(\ref{eq:IvsrhoE}), (\ref{eq:IvsrhoEII}) 
represent the constraints on the variations of 
$\bar{\rho}_{\rm man}$, and $\bar{\rho}_{\rm OC}$, $\bar{\rho}_{\rm IC}$ 
or $\bar{\rho}_{\rm C}$, from the knowledge of $M_\oplus$ and $I_\oplus$.

 We will assume that 
$R_{\rm man}$, $R_{\rm OC}$ and $R_{\rm IC}$ 
and, correspondingly, the values of the 
constants $r_{\rm man}$,  $r_{\rm OC}$ and  $r_{\rm IC}$ 
are correctly given by PREM. 
These are standard assumptions 
made in the neutrino tomography of the Earth 
studies of possible deviations of 
$\bar{\rho}_{\rm man}$, $\bar{\rho}_{\rm OC}$ and $\bar{\rho}_{\rm IC}$
from their PREM values. 
In our further analyses we will consider 
the ``surface'' terms in Eqs. (\ref{eq:rhoEII}) and 
(\ref{eq:IvsrhoE}), $a_{\rm oc1c2}\,\bar{\rho}_{\rm oc1c2}$ 
and $b_{\rm oc1c2}\,\bar{\rho}_{\rm oc1c2}$, as fixed.
Under these conditions,  Eqs. (\ref{eq:rhoEII}) and (\ref{eq:IvsrhoE})
represent two equations for the three densities of interest: 
$\bar{\rho}_{\rm man}$, $\bar{\rho}_{\rm OC}$ and $\bar{\rho}_{\rm IC}$.

\vspace{0.3cm}
%
\section{Implementing the $M_{\oplus}$, $I_{\oplus}$ and EHE  
Constraints}
\label{sec:implMI}
%
%
\vspace{0.3cm}
%
\subsection{
The Mantle, Outer Core and Inner Core Case
}
\label{sec:implMI3layers}
%

 Suppose one is interested in  
obtaining information about possible deviations of the  
densities of the mantle $\rho_{\rm man}$, the   
outer core, $\rho_{\rm OC}$, and the inner core, $\rho_{\rm IC}$
from their PREM  values. 
In the corresponding statistical analysis one varies the density 
in the chosen layer of interest,  
say $\rho_{\rm OC}$. In order to satisfy the $M_\oplus$ and  $I_{\oplus}$ 
constraints one has to compensate the variation of 
$\rho_{\rm OC}$ 
by an appropriate change of the densities in the other two layers, 
$\rho_{\rm man}$ and $\rho_{\rm IC}$
In the analyses performed so far it is assumed that the indicated 
variations and changes of interest are by constant $r$-independent 
factors $\kappa_i$, which can be positive, negative or zero:
\begin{equation}
\rho^\prime_{i} = (1 + \kappa_i)\,\rho_{i}\,,~i= man, OC,IC\,. 
\label{eq:kappai}
\end{equation}
%
We will analyse this possibility below.

The total Earth mass and momentum of inertia constraints 
in the simple case we are considering of average densities 
imply that $\bar{\rho}^\prime_{i}$ should satisfy 
Eqs. (\ref{eq:rhoEII}) and (\ref{eq:IvsrhoE}) 
and the following two equations:
\begin{eqnarray}
\nonumber
& a_{\rm man}\, (1 + \kappa_{\rm man})
\bar{\rho}_{\rm man} + a_{\rm OC}\,(1 + \kappa_{\rm OC})\bar{\rho}_{\rm OC} 
+ a_{\rm IC}\,(1 + \kappa_{\rm IC})\bar{\rho}_{\rm IC} 
+ a_{\rm oc1c2}\,\bar{\rho}_{\rm oc1c2}
= \bar{\rho}_{\oplus}\,,\\[0.25cm]
& b_{\rm man}\,(1 + \kappa_{\rm man}) \bar{\rho}_{\rm man} + 
 b_{\rm OC}\,(1 + \kappa_{\rm OC})\bar{\rho}_{\rm OC} +
b_{\rm IC}\,(1 + \kappa_{\rm IC})\bar{\rho}_{\rm IC} 
+ b_{\rm oc1c2}\,\bar{\rho}_{\rm oc1c2}
=  C^\prime\,\bar{\rho}_\oplus\,.
\label{eq:rhoprime}
\end{eqnarray}
%
Combining these equations with Eqs. (\ref{eq:rhoEII}) and 
(\ref{eq:IvsrhoE}) and taking into account that the terms 
$a_{\rm oc1c2}\,\bar{\rho}_{\rm oc1c2}$ 
and $b_{\rm oc1c2}\,\bar{\rho}_{\rm oc1c2}$
are fixed, we get neglecting the uncertainties 
in the determination of $M_\oplus$ and $I_\oplus$:
\begin{eqnarray}
\nonumber
& a_{\rm man}\,\kappa_{\rm man}
\bar{\rho}_{\rm man} + a_{\rm OC}\,\kappa_{\rm OC}
\bar{\rho}_{\rm OC} 
+ a_{\rm IC}\,\kappa_{\rm IC}
\bar{\rho}_{\rm IC} = 0\,,\\[0.25cm]
& b_{\rm man}\,\kappa_{\rm man} \bar{\rho}_{\rm man} + 
 b_{\rm OC}\,\kappa_{\rm OC}
\bar{\rho}_{\rm OC} +
b_{\rm IC}\,\kappa_{\rm IC}
\bar{\rho}_{\rm IC} = 0\,.
 \label{eq:eqkappai2}
\end{eqnarray}
%

Suppose we vary $\rho_{\rm OC}$ and are interested 
in the sensitivity of a given detector 
to possible deviations of $\rho^\prime_{\rm OC}$ from 
the PREM prediction, i.e., to values of 
$\kappa_{\rm OC} \neq 0$.
In the simple case of average densities we are considering 
we can express $\kappa_{\rm man}$ and $\kappa_{\rm IC}$ 
in terms of $\kappa_{\rm OC}$ using Eq. (\ref{eq:eqkappai2}):
\begin{eqnarray}
\nonumber
& \kappa_{\rm man} = \dfrac{\kappa_{\rm OC}}{D_{OC}}\, 
\left(a^\prime_{\rm IC}\,b_{\rm OC} - a_{\rm OC}\,b^\prime_{\rm IC} \right)\,,
\\[0.25cm] 
& \kappa_{\rm IC} = \dfrac{\kappa_{\rm OC}}{D_{OC}}\, 
\left(a_{\rm OC}\,b^\prime_{\rm man} - a^\prime_{\rm man}\,b_{\rm OC} \right)\,.
 \label{eq:kappaICman}
\end{eqnarray}
%
Here 
\begin{equation}
D_{OC} \equiv a^\prime_{\rm man}\,b^\prime_{\rm IC} -\, 
   a^\prime_{\rm IC}\,b^\prime_{\rm man}\,,
\label{eq:DOC}   
\end{equation}
%
and
\begin{equation}
 a^\prime_{j} \equiv \dfrac{\bar{\rho}_j}{\bar{\rho}_{\rm OC}}\,a_j\,,~~
b^\prime_{j} \equiv \dfrac{\bar{\rho}_j}{\bar{\rho}_{\rm OC}}\,b_j\,,~~
j=man,IC\,.
\label{eq:apbp}
\end{equation}
%
It follows, in particular, from Eq. (\ref{eq:kappaICman}) 
that in order to satisfy the Earth total mass and moment of inertia 
constraints, the variations of  $\bar{\rho}_{\rm OC}$ have to be compensated 
by changes of the densities of at least two other Earth layers. 
Using the PREM values of $\bar{\rho}_{\rm IC}$, $\bar{\rho}_{\rm OC}$
and  $\bar{\rho}_{\rm man}$ 
given in Eqs. (\ref{eq:rhoICIIC}), (\ref{eq:rhoOCIOC}) and (\ref{eq:rhomIm}), 
as well as the values of $a_i$ 
(in Eq. (\ref{eq:a})) and $b_i$ (in Eq. (\ref{eq:b})) we find:
\begin{equation}
\kappa_{\rm man} \cong -\,0.1104\,\kappa_{\rm OC}\,,~~
\kappa_{\rm IC} \cong  -\,13.9558\,\kappa_{\rm OC}\,.
\label{eq:kmanIC2}
\end{equation}
%

In the case of $\kappa_{\rm OC} > 0$ the relevant Earth hydrostatic equilibrium 
(EHE)  constraint is 
$\bar{\rho}_{\rm OC} (1 + \kappa_{\rm OC}) 
< \bar{\rho}_{\rm IC} (1 + \kappa_{\rm IC})$, with  $\kappa_{\rm IC}<0$.
Using the relation between $\kappa_{\rm IC}$ and $\kappa_{\rm OC}$
given in Eq. (\ref{eq:kmanIC2}), we find:
\begin{equation}
\kappa_{\rm OC} < \dfrac{\bar{\rho}_{\rm IC} - \bar{\rho}_{\rm OC}}
{\bar{\rho}_{\rm OC} + 13.9558\bar{\rho}_{\rm IC}} \cong 0.0104\,, 
\label{eq:rhoOC31}
\end{equation}
%
where we have used the PREM values of $\bar{\rho}_{\rm IC}$ 
and $\bar{\rho}_{\rm OC}$ (see Eq. (\ref{eq:rhoICIIC})).
This result implies the following lower limits:
$\kappa_{\rm man} > -\,1.15\times 10^{-3}$  
and $\kappa_{\rm IC} >  -\, 0.145$. 

For  $\kappa_{\rm OC} < 0$ the relevant EHE constraint is 
instead $\bar{\rho}_{\rm man} (1 + |\kappa_{\rm man}|) 
< \bar{\rho}_{\rm OC} (1 - |\kappa_{\rm OC}|)$. 
This inequality together with the relation between 
$\kappa_{\rm man}$ and $\kappa_{\rm OC}$ 
we have derived and the PREM values of $\bar{\rho}_{\rm man}$ 
and $\bar{\rho}_{\rm OC}$ leads to the following lower limit: 
\begin{equation}
\kappa_{\rm OC} > -\, \dfrac{\bar{\rho}_{\rm OC} - \bar{\rho}_{\rm man}}
{\bar{\rho}_{\rm OC} + 0.1104\bar{\rho}_{\rm man}} \cong -\, 0.546\,,~~
\kappa_{\rm OC} < 0\,.
\label{eq:rhoOC32}
\end{equation}
%
This lower limit on $\kappa_{\rm OC} < 0$ corresponds to the following 
upper limits on $\kappa_{\rm man}$ 
and $\kappa_{\rm IC}$: $\kappa_{\rm man} < 0.060$ 
and $\kappa_{\rm IC} < 7.62$.   
Summarising we have:
\begin{align}
& -\,0.546 < \kappa_{\rm OC} < 0.010\,,
\nonumber \\
&   -\,1.15\times 10^{-3} < \kappa_{\rm man} <  0.060\,,
\nonumber \\
& -\,0.145 < \kappa_{\rm IC} < 7.62\,.
\label{eq:ka3l1a}
\end{align}
%
In terms of average densities these limits imply:
\begin{align}
& 4.950\,\dfrac{g}{cm^3}< \bar{\rho}_{\rm OC} < 11.010\,\dfrac{g}{cm^3}\,,
\nonumber \\
& 4.660\,\dfrac{g}{cm^3} < \bar{\rho}_{\rm man} < 4.945\,\dfrac{g}{cm^3} \,,
\nonumber \\
& 11.024\,\dfrac{g}{cm^3} < \bar{\rho}_{\rm IC} < 111.143\,\dfrac{g}{cm^3} \,.
\label{eq:rho3lOCICman}
\end{align}
%

We see that the $M_{\oplus}$, $I_{\oplus}$ and EHE conditions  
severely constrain the possible positive deviations of $\bar{\rho}_{\rm OC}$, 
and both positive and negative deviations $\bar{\rho}_{\rm man}$, 
from their respective PREM values.
At the same time the obtained lower limit on $\kappa_{\rm OC} < 0$
allows excessively large values of $\kappa_{\rm IC} > 0$ 
and thus of $\bar{\rho}_{\rm IC} (1 + \kappa_{\rm IC})$,
as large as,  e.g., 8.6 times bigger than the PREM value of  
$\bar{\rho}_{\rm IC}$. 
The existing seismological data on $\bar{\rho}_{\rm IC}$ 
is claimed to have an uncertainty not exceeding $\sim 10\%$.
Even if we allow a rather large maximum deviation of  $\bar{\rho}_{\rm IC}$ from 
its PREM value of 25\% (50\%), which corresponds to 
$\kappa_{\rm IC} \leq 0.25~(0.50)$, this leads to the following lower limit 
on $\kappa_{\rm OC} < 0$: $\kappa_{\rm OC} \gtap -\,0.018~(0.036)$.

Thus, we see that in the considered case of three layer Earth 
density distribution, the Earth mass and moment of  inertia constraints 
together with the requirement of the Earth being in hydrostatic equilibrium 
and the value of the inner core average density lead to severe 
constraints on the possible uniform deviations of $\bar{\rho}_{\rm OC}$ 
from its PREM value: 
\begin{equation}
-\,0.018~(0.036) \ltap \kappa_{\rm OC} < 0.010\,.
\label{eq:kappaOCmanICOC}
\end{equation}
%
These very restrictive constraints can be traced to 
the relatively small contributions of the
inner core to $M_\oplus$ and especially to $I_\oplus$.

We note that we get practically the same results 
if we consider the ``surface'' layers 
as part of the mantle. In this case the mantle 
extends from $R_{\rm C} = 3480$ km to $R_\oplus = 6371$ km 
(instead of $R_{\rm man} = 6346.6$ km). 

 In this analysis we did not take into account the uncertainties 
in the determination of $M_\oplus$ and $I_\oplus$. 
As it follows from Eqs. (\ref{eq:ME}) and  (\ref{eq:IE}), 
the relative $1\sigma$ uncertainties in the values of 
$M_\oplus$ and $I_\oplus$ are $\sim 10^{-4}$. 
These uncertainties are too small to have any significant effect, 
if taken into account, on the results we have obtained.

\vspace{0.3cm}
%
\subsection{The case of Two Mantle Layers,  Outer Core and Inner Core
}
\label{sec:implMI4layers}
%

We consider next the case of two different mantle layers 
and the already discussed outer core, inner core 
and ``surface'' layers. The mantle is divided into 
i)  lower mantle, extending from $R_c = 3480$ km to $R_{\rm lman} = 5701$ km,  
with densities $\bar{\rho}_{\rm lman}$ and $\bar{\rho}^{\rm I}_{\rm lman}$, and 
ii) upper mantle, extending from  $R_{\rm lman} = 5701$ km to 
$R_{\rm man} = 6346.6$ km, 
with   densities $\bar{\rho}_{\rm uman}$  and $\bar{\rho}^{\rm I}_{\rm uman}$.
It is not difficult to calculate the PREM values of 
$\bar{\rho}_{\rm lman}$, $\bar{\rho}^{\rm I}_{\rm lman}$,
$\bar{\rho}_{\rm uman}$  and $\bar{\rho}^{\rm I}_{\rm uman}$:
\begin{eqnarray}
& \bar{\rho}_{\rm lman} = 4.9035~\dfrac{g}{cm^3}\,,~~
\bar{\rho}^{\rm I}_{\rm lman} = r_{\rm lman} \bar{\rho}_{\rm lman}\,,~~ 
 r_{\rm lman} = 0.9771\,, 
\label{eq:rhorlman}
\\[0.20cm]
& \bar{\rho}_{\rm uman} = 3.6046~\dfrac{g}{cm^3}\,,~~ 
\bar{\rho}^{\rm I}_{\rm man} = r_{\rm lman} \bar{\rho}_{\rm lman}\,,~~
 r_{\rm uman} = 0.9425\,. 
\label{eq:rhoruman}
\end{eqnarray}
%
As in the preceding subsection we will assume that 
that the  variations and changes of 
$\bar{\rho}_{\rm IC}$, $\bar{\rho}_{\rm OC}$, 
 $\bar{\rho}_{\rm lman}$ and   $\bar{\rho}_{\rm uman}$ 
are by constant $r$-independent 
factors $\kappa_{\rm IC}$, $\kappa_{\rm OC}$,  $\kappa_{\rm lman}$
and  $\kappa_{\rm uman}$:
\begin{equation}
\rho^\prime_{i} = (1 + \kappa_i)\,\rho_{i}\,,~i={\rm IC,OC,lman,uman}\,. 
\label{eq:kappai4}
\end{equation}
%
The analogs of the equations in Eq. (\ref{eq:eqkappai2}) now read:
\begin{eqnarray}
\nonumber
& a_{\rm lman}\,\kappa_{\rm lman} \bar{\rho}_{\rm lman} 
+ a_{\rm uman}\,\kappa_{\rm uman} \bar{\rho}_{\rm uman} 
+ a_{\rm OC}\,\kappa_{\rm OC}\bar{\rho}_{\rm OC} 
+ a_{\rm IC}\,\kappa_{\rm IC}
\bar{\rho}_{\rm IC} = 0\,,\\[0.25cm]
& b_{\rm lman}\,\kappa_{\rm lman} \bar{\rho}_{\rm lman} +
b_{\rm uman}\,\kappa_{\rm uman} \bar{\rho}_{\rm uman}
+ b_{\rm OC}\,\kappa_{\rm OC}\bar{\rho}_{\rm OC} +
b_{\rm IC}\,\kappa_{\rm IC} \bar{\rho}_{\rm IC} = 0\,.
\label{eq:eqkappaiII}
\end{eqnarray}
%

For the new parameters in these equations we find:
\begin{eqnarray}
&a_{\rm lman} = 
\dfrac{R^3_{\rm lman}}{R^3_\oplus}(1 - \dfrac{R^3_{\rm C}}{R^3_{\rm lman}}) 
\cong 0.5535\,,~ 
a_{\rm uman} = \dfrac{R^3_{\rm man}}{R^3_\oplus} 
\left(1 - \dfrac{R^3_{\rm lman}}{R^3_{\rm man}}\right) 
 \cong 0.2720\,,   
\label{eq:aluman}
 \\[0.20cm] 
&b_{\rm lman} = r_{\rm lman} \dfrac{R^5_{\rm lman}}{R^5_\oplus}
\left (1 - \dfrac{R^5_{\rm c}}{R^5_{\rm lman}}\right)
\cong 0.5131\,,~
b_{\rm uman} = r_{\rm uman}\dfrac{R^5_{\rm man}}{R^5_\oplus}
\left (1 - \dfrac{R^5_{\rm lman}}{R^5_{\rm man}}\right)
\cong 0.3838\,. 
\label{eq:bluman}
\end{eqnarray}
%
where we have used the values of
$R_{\rm C}$, $R_{\rm lman}$, $R_{\rm man}$, $R_\oplus$, 
$r_{\rm lman}$ and  $r_{\rm uman}$. 

Equations (\ref{eq:eqkappaiII}) represent the 
constraints following from the knowledge of 
$M_\oplus$ and $I_\oplus$ on the  the possible 
deviations of $\bar{\rho}_{\rm IC}$, $\bar{\rho}_{\rm OC}$, 
$\bar{\rho}_{\rm lman}$ and   $\bar{\rho}_{\rm uman}$ 
from their PREM values.

Suppose we vary $\rho_{\rm OC}$ and are interested 
in the sensitivity of a given detector 
to possible deviations $\kappa_{\rm OC} \neq 0$ 
of the PREM prediction for $\rho_{\rm OC}$. 
Consider the simple case of $\kappa_{\rm IC} = 0$. 
Working with the  average densities
and using Eq. (\ref{eq:eqkappaiII}) 
we can express $\kappa_{\rm lman}$ and $\kappa_{\rm uman}$ 
in terms of $\kappa_{\rm OC}$:
\begin{eqnarray}
\nonumber
& \kappa_{\rm lman} = \dfrac{\kappa_{\rm OC}}{D^{(m)}_{OC}}\, 
\left(a^\prime_{\rm uman}\,b_{\rm OC} - a_{\rm OC}\,b^\prime_{\rm uman} \right)\,,
\\[0.25cm] 
& \kappa_{\rm uman} = \dfrac{\kappa_{\rm OC}}{D^{(m)}_{OC}}\, 
\left(a_{\rm OC}\,b^\prime_{\rm lman} - a^\prime_{\rm lman}\,b_{\rm OC} \right)\,.
 \label{eq:kappaICman4}
\end{eqnarray}
%
Here 
\begin{equation}
D^{(m)}_{OC} \equiv a^\prime_{\rm lman}\,b^\prime_{\rm uman} -\, 
   a^\prime_{\rm uman}\,b^\prime_{\rm lman}\,,
\label{eq:DOC4}   
\end{equation}
%
and
\begin{equation}
 a^\prime_{j} \equiv \dfrac{\bar{\rho}_j}{\bar{\rho}_{\rm OC}}\,a_j\,,~~
b^\prime_{j} \equiv \dfrac{\bar{\rho}_j}{\bar{\rho}_{\rm OC}}\,b_j\,,~~
 j=lman,uman\,.
\label{eq:apbp4}
\end{equation}
%
Using the PREM values of $\bar{\rho}_{\rm OC}$, $\bar{\rho}_{\rm lman}$
and  $\bar{\rho}_{\rm uman}$ given in Eqs. (\ref{eq:rhoOCIOC}) and 
(\ref{eq:rhorlman}) and (\ref{eq:rhoruman}), 
as well as the values of $a_i$ and $b_i$, $i=OC,lman,uman$,
from  Eqs. (\ref{eq:a}), (\ref{eq:bmOCICnum}),
(\ref{eq:aluman})  and (\ref{eq:bluman}), we get:
\begin{equation}
\kappa_{\rm lman}  = f_{\rm lman}\,\kappa_{\rm OC}\,,~
f_{\rm lman} \cong -\,1.4327;~~~
\kappa_{\rm uman} = f_{\rm uman}\,\kappa_{\rm OC}\,,~
f_{\rm uman}\cong 2.2329. 
\label{eq:kluman}
\end{equation}
%
Thus, in order to compensate the variation of the density in the 
outer core, the density should be increased in one of the mantle 
layers and decreased in the second mantle layer.

  The one of the EHE constraints  
for $\kappa_{\rm OC} > 0$ is 
$\bar{\rho}_{\rm OC} (1 + \kappa_{\rm OC}) 
< \bar{\rho}_{\rm IC}$ (recall that we have set $\kappa_{\rm IC}=0$).
Using the PREM values of $\bar{\rho}_{\rm IC}$ 
and $\bar{\rho}_{\rm OC}$ (see Eq. (\ref{eq:rhoICIIC})) we find:
\begin{equation}
\kappa_{\rm OC} < \dfrac{\bar{\rho}_{\rm IC} - \bar{\rho}_{\rm OC}}
{\bar{\rho}_{\rm OC}} \cong 0.18\,. 
\label{eq:EHEul1}
\end{equation}
%
In the case we are considering the following EHE conditions  
should also be satisfied:
\begin{equation}
\bar{\rho}_{\rm uman} < \bar{\rho}_{\rm lman} < \bar{\rho}_{\rm OC}\,.
\label{eq:EHE2}
\end{equation}
%
The strongest upper limit on $\kappa_{\rm OC} > 0$
follows from the EHE requirement: 
$\bar{\rho}_{uman}(1 + \kappa_{uman}) < \bar{\rho}_{lman}(1 + \kappa_{lman})$:
\begin{equation}
\kappa_{\rm OC} < \dfrac{\bar{\rho}_{\rm lman} - \bar{\rho}_{\rm uman}}
{\bar{\rho}_{\rm uman}f_{uman} - \bar{\rho}_{\rm lman}f_{lman}} \cong 0.086\,. 
\label{eq:EHEul2}
\end{equation}
%
This upper limit on  $\kappa_{\rm OC}$ is approximately by  a factor of two
smaller than the upper limit quoted in Eq. (\ref{eq:EHEul1}).

 In the case of  $\kappa_{\rm OC} < 0$, the EHE constraint reads:
$\bar{\rho}'_{\rm man} < \bar{\rho}_{\rm OC} (1 - |\kappa_{\rm OC}|)$,
where  $\bar{\rho}'_{\rm man}$ is the modified mantle density 
due to the changes of the lower and upper mantle densities 
$\bar{\rho}_{\rm lman}(1 + \kappa_{\rm lman})$ and   
$\bar{\rho}_{\rm uman}(1 +  \kappa_{\rm uman})$.
It is not difficult to show that
\begin{equation}
\bar{\rho}'_{\rm man} = \bar{\rho}_{\rm lman}(1 + \kappa_{\rm lman})\,C_{\rm lman} 
+  \bar{\rho}_{\rm uman}(1 + \kappa_{\rm uman})\,C_{\rm uman}\,.
\label{eq:rhopman}
\end{equation}
%
Here 
\begin{equation}
C_{\rm lman} = \dfrac{R^3_{\rm lman}}{R^3_{\rm man}} 
\left(1 - \dfrac{R^3_{\rm C}}{R^3_{\rm man}}\right)\cong 0.670\,,~~
C_{\rm uman} = 
\dfrac{1 - \dfrac{R^3_{\rm lman}}{R^3_{\rm man}}}
{1 - \dfrac{R^3_{\rm C}}{R^3_{\rm man}}}\cong 0.329\,,
\label{eq:Cluman}
\end{equation}
%
where we have used the values of $R_{\rm lman}$, $R_{\rm man}$ and $R_{\rm C}$.
The EHE constraint in the considered case has the form:
\begin{equation}
\kappa_{\rm OC} > -\, 
\dfrac{\bar{\rho}_{\rm OC} - \bar{\rho}_{\rm lman}\,C_{\rm lman} - 
\bar{\rho}_{\rm uman}\,C_{\rm uman}}
{\bar{\rho}_{\rm OC} - \bar{\rho}_{\rm lman}\,C_{\rm lman}\,f_{\rm lman} -  
\bar{\rho}_{\rm uman}\,C_{\rm uman}\,f_{\rm uman}
}\,,~~\kappa_{\rm OC} < 0\,.
\label{eq:EHE2a}
\end{equation}
%
Using the values of $\bar{\rho}_{\rm OC}$, $\bar{\rho}_{\rm lman}$, 
$\bar{\rho}_{\rm uman}$, $f_{\rm lman}$, $f_{\rm uman}$, 
$C_{\rm lman}$ and $C_{\rm uman}$ given in 
Eqs. (\ref{eq:rhoOCIOC}), (\ref{eq:rhorlman}), 
(\ref{eq:rhoruman}), (\ref{eq:kluman}) and (\ref{eq:Cluman})
we get 
\footnote{We note that in the expression in the denominator in the right-hand 
side of Eq. (69) in the published version of this article, 
Ref. \cite{Petcov:2024icq}, the 
signs in front of the 2nd and 3rd terms 
should be negative, as in Eq. (\ref{eq:EHE2a}) above.
Correspondingly, the lower limit on $\kappa_{\rm OC}$ in Eqs. (70) and 
(71) in \cite{Petcov:2024icq} 
changes to $\kappa_{\rm OC} > -\,0.496$ (Eq. (\ref{eq:EHE2b})).
}:
\begin{equation}
\kappa_{\rm OC} > -\,0.496\,.
\label{eq:EHE2b}
\end{equation}
%
However, the physical requirement $\kappa_{\rm uman} > -\,1$ 
leads to a somewhat stronger limit of $\kappa_{\rm OC}$: 
$\kappa_{\rm OC} > -\,0.448$.
%
We have to ensure also that the EHE constraint 
in Eq. (\ref{eq:EHE2}),  
$\bar{\rho}_{\rm lman}(1 + \kappa_{\rm lman})  < 
\bar{\rho}_{\rm OC}(1 + \kappa_{\rm OC})$,  
is fulfilled. In the relevant case of 
$\kappa_{\rm OC} < 0$ this leads to the strongest lower limit on 
$\kappa_{\rm OC}$:
\begin{equation}
\kappa_{\rm OC} > -\,0.334\,. 
\label{eq:EHE3c}
\end{equation}
%

Combining the limits we have obtained on $\kappa_{\rm OC}$ we have 
\footnote{The EHE constraints in Eq. (\ref{eq:EHE2}) were not taken 
into account in \cite{Petcov:2024icq}. Taking them into account leads to the 
inequalities given in Eqs. (\ref{eq:EHEul2}) and (\ref{eq:EHE3c}), and thus 
to those in Eq. (\ref{eq:kappaOClu}), which are more restrictive
and correct the constraints on  $\kappa_{\rm OC}$ reported in Eq. (71)  
in \cite{Petcov:2024icq}.
}:
\begin{equation}
-\,0.334 < \kappa_{\rm OC} < 0.086\,.
\label{eq:kappaOClu}
\end{equation}
%
Using the relations in Eq. (\ref{eq:kluman}) we get the 
corresponding constraints on  $\kappa_{\rm lman}$ and $\kappa_{\rm uman}$:
\begin{align}
 &-\,0.123 < \kappa_{\rm lman} < 0.479 \nonumber\\
 &-\,0.747 < \kappa_{\rm uman} < 0.192\,.
 \label{eq:ka3lmum}     
\end{align}
%
These results limit the ranges of the possible values of 
$\bar{\rho}_{\rm OC}$, $\bar{\rho}_{\rm lman}$, 
$\bar{\rho}_{\rm uman}$ to:
\begin{align}
 & 7.25\,\dfrac{g}{cm^3} < \bar{\rho}_{\rm OC} < 11.84\,\dfrac{g}{cm^3}\,,
\nonumber\\
 & 4.30\, \dfrac{g}{cm^3} < \bar{\rho}_{\rm lman} < 7.25\,\dfrac{g}{cm^3}\,,
\nonumber \\
 & 0.912\,\dfrac{g}{cm^3} < \bar{\rho}_{\rm uman} < 4.25\,\dfrac{g}{cm^3}\,.
\label{eq:ka3lrho}     
\end{align}
%

Practically  the same results are obtained
if we include the ``surface'' layers 
in the upper mantle. In this case 
the ``surface'' layers and the upper mantle form 
one ``upper mantle'' layer, which 
extends from $R_{\rm lman} = 3701$ km to $R_\oplus = 6371$ km 
(instead of $R_{\rm man} = 6346.6$ km). 
Similarly, the results change insignificantly if 
one joins the inner core and the outer core 
in one core layer.

Thus, we see that in the case 
of two mantle, outer and inner core (or one core) layers, 
the $M_\oplus$, $I_\oplus$ and EHE 
constraints on the possible variation of 
$\bar{\rho}_{\rm OC}$ are much less restrictive 
than in the case of one mantle, outer and inner core layers,  
considered in the preceding subsection.
The main reason for this difference is that 
the contributions of the inner core 
to  $M_\oplus$ and $I_\oplus$ are not 
crucial for satisfying the $M_\oplus$ and $I_\oplus$ 
constraints when there are two mantle 
layers, while, being relatively small, 
they are critical in the one mantle 
layer case.

 One may consider other layer structures of the Earth density distribution, 
for example, inner core, two equal width outer core layers and 
mantle layer. We did not performed the numerical analysis in this case 
but expect the $M_\oplus$, $I_\oplus$ and EHE 
constraints to be relevant but much less restrictive than in the case of 
the three layer case we have considered.

\section{Summary and Conclusions}
\label{sec:summ}

Summarising, we have shown that the impact of the 
Earth total mass, $M_\oplus$, moment of inertia, $I_\oplus$, and 
the Earth hydrostatic equilibrium (EHE) constraints 
in the neutrino tomography studies of the Earth density structure 
depends crucially on the Earth layer structure assumed in the studies.
More specifically, we find that if one approximates the Earth density 
distribution with three layers - inner core, outer core and mantle, 
the $M_\oplus$, $I_\oplus$ and EHE constraints lead to severe limits 
on the possible uniform deviations of the outer core and mantle
densities from their PREM values even if one assumes that the inner core 
density, which plays a crucial role in satisfying the constraints, 
can be by 50\% larger than its PREM density.
We have shown also that, in contrast, the $M_\oplus$, 
$I_\oplus$ and EHE constraints on the possible deviations 
of the  outer core and mantle densities from their PREM value are much less
restrictive if one assumes a four layer Earth density structure 
consisting of  inner core, outer core, and two - upper and lower - 
mantle layers. In this case the  inner core 
density plays an insignificant (or no) role in satisfying 
the constraints. 
  
  It follows from our results  that the implementation of  $M_\oplus$, 
$I_\oplus$ and EHE constraints in the neutrino tomography studies of 
the Earth density structure can have a very strong impact on the 
conclusions reached in the studies.

\section*{Acknowledgements}
  
 This work was supported in part by the European 
Union's Horizon 2020 research and innovation programme under 
the Marie 
Sklodowska-Curie grant agreement No.~860881-HIDDeN, 
by the INFN program on Theoretical Astroparticle Physics 
and by the  World Premier International Research Center
Initiative (WPI Initiative, MEXT), Japan.

\end{document}